\documentclass[fleqn,usenatbib,useAMS]{mnras}

\usepackage[dvipsnames]{xcolor}
\usepackage{tikz,hyperref}

\definecolor{lime}{HTML}{A6CE39}
\DeclareRobustCommand{\orcidicon}{
	\begin{tikzpicture}
	\draw[lime, fill=lime] (0,0) 
	circle [radius=0.13] 
	node[white] {{\fontfamily{qag}\selectfont \tiny ID}};
	\draw[white, fill=white] (-0.0625,0.095) 
	circle [radius=0.007];
	\end{tikzpicture}
	\hspace{-2mm}
}

\foreach \x in {A, ..., Z}{\expandafter\xdef\csname orcid\x\endcsname{\noexpand\href{https://orcid.org/\csname orcidauthor\x\endcsname}
			{\noexpand\orcidicon}}
}

\usepackage{newtxtext,newtxmath}

\usepackage{booktabs}

\usepackage[T1]{fontenc}
\usepackage{ae,aecompl}

\usepackage{graphicx}	
\usepackage{amsmath}	
\usepackage{booktabs}
\usepackage{times}
\input{epsf}

\title[The WR phenomenon through low-mass stars]{Peering into the Wolf-Rayet phenomenon through [WO] and [WC] stars}

\author[J.~A.~Toal\'{a}, H.~Todt \& A.~A.~C.~Sander]{Jes\'{u}s~A.~Toal\'{a}\thanks{E-mail:\,j.toala@irya.unam.mx}$^{1\orcidA}$,  Helge~Todt$^{2\orcidB}$ and Andreas~A.~C.~Sander$^{3\orcidC}$
\\
$^{1}$Instituto de Radioastronom\'{i}a y Astrof\'{i}sica, Universidad Nacional Aut\'{o}noma de M\'{e}xico, 58090 Morelia, Michoac\'{a}n, Mexico\\
$^{2}$Institut f\"{u}r Physik und Astronomie, Universit\"{a}t Potsdam, Karl-Liebknecht-Str. 24/25, 14476 Potsdam, Germany\\
$^{3}$Zentrum f\"{u}r Astronomie der Universit\"{a}t Heidelberg, Astronomisches Rechen-Institut, M\"{o}nchhofstr. 12-1469120 Heidelberg, Germany
}

\date{\today}

\pubyear{2023}

\begin{document}
\label{firstpage}
\pagerange{\pageref{firstpage}--\pageref{lastpage}}
\maketitle

\begin{abstract}
Spectroscopic observations have shown for decades that the Wolf-Rayet (WR) phenomenon is ubiquitous among stars with different initial masses. Although much effort to understand the winds from massive WR stars has been presented in the literature, not much has been done for such type of stars in the low-mass range. Here we present an attempt to understand the winds from [WR]-type stars using results from spectral analyses with the full non-LTE stellar atmosphere code {\scshape PoWR}. These results are put into context with the properties of massive WR stars. We found that WC+[WC] stars and WO+[WO] stars create independent sequences in the mass-loss rate ($\dot{M}$) and modified wind momentum ($D_\mathrm{mom}$) versus luminosity ($L$) diagrams. Our analysis indicates that even when the winds of WR and [WR] stars become optically thin, there is no breakdown of the general mass-loss trend, contrary to the observed ``weak wind phenomenon'' in OB stars. We report that all WR-type stars studied here broadly define single sequences in the wind efficiency ($\eta$) versus transformed mass-loss rate ($\dot{M}_\mathrm{t}$), the $\dot{M}_\mathrm{t}$--$T_\mathrm{eff}$ diagram, and the $(L, T_\mathrm{eff}, \dot{M})$ space, which suggest these to be fundamental properties of the WR phenomenon (regardless of the mass range), at least for WR-type stars of the O and C sequences. Our analytical estimations could drive computations of future stellar evolution models for WR-type stars.
\end{abstract}

\begin{keywords}
stars: low-mass stars --- stars: winds, outflows --- stars: Wolf–Rayet --- stars: evolution --- (ISM:) planetary nebulae   
\end{keywords}



\section{Introduction}\label{introduction}
\label{sec:intro}

Classical Wolf-Rayet (WR) stars are the descendants of massive ($M_\mathrm{i}>20$ M$_\odot$) stars that loose their H-rich envelops as a consequence of their powerful winds during previous evolutionary stages \citep{Conti1975} or as a result of binary interactions \citep{Paczynski1967}. A more recent scenario in which strong internal mixing efficiency creates He-rich, non-stripped stars has also been suggested \citep{Hainich2015,Brott2011}. Regardless of their formation scenario, they are characterised by the presence of broad emission lines of He, C, N and O in their spectra \citep[e.g.,][]{Crowther2007} and, depending on the strength of the different emission lines, they can be classified as nitrogen (WN), carbon (WC) or oxygen types (WO).

Observations and stellar atmosphere models have demonstrated that WR stars posses the most powerful winds among massive hot stars. The classic CAK theory applied for OB stars \citep{Castor1975} including its later extensions \citep[e.g,][]{Pauldrach1986,Kudritzki1999} fail in the WR regime. Beside exceeding the single-scattering limit \citep{LucyAbbott1993,Springmann1994}, in particular classical, i.e. helium-burning, WR stars can have a complex, multi-bump opacity structure \citep[e.g.,][]{NugisLamers2002,Grafener2005,Sander2020a} that defy any CAK-type parametrization of the radiative force. Previous authors have shown that OB stars are consistently located below the WR stars in the $\dot{M}$--$L$ diagram \citep[e.g.,][]{Vink2000,Vink2017,Ramachandran2019}. Moreover, there seems to be a breakdown of the mass-loss rates at $\log_{10}$($L$/L$_\odot$)$\lesssim$5 at Galactic metallicity for late O and early B stars. This is known as the ``weak wind problem'' \citep[e.g.,][]{Martins2005,Marcolino2009,Oskinova2011,Huenemoerder2012}. These empirical findings as well as recent theoretical studies \citep[e.g.,][]{Sander2020b,Sander2023,Bjoerklund2023,Krticka2024} indicate that the underlying wind-driving process behaves differently for massive OB-type stars and WR stars.

\begin{table*}
\begin{center}
\caption{Stellar and wind properties of the [WR] stars used in this work.}
\setlength{\tabcolsep}{0.9\tabcolsep}  
\label{tab:stars}
\begin{tabular}{lccccccccl} 
\hline
Name      & Spec. Type &  log($L$)  & log($\dot{M}$)       & $T_\mathrm{eff}$($\tau$=2/3) & $R_\star$($\tau$=2/3)  & $\varv_\infty$   & log($D_\mathrm{mom}$)     & $\eta$ & Reference\\
          &            & (L$_\odot$)& (M$_\odot$~yr$^{-1}$)& (kK) &(R$_\odot$)& (km~s$^{-1}$)& (g~cm~s$^{-2}$~R$_\odot^{1/2}$) &  & \\
\hline
NGC\,2371 & [WO1] & 3.45 & $-$7.75  & 130 & 0.105 & 3700 & 26.13 & 1.11 & \citet{GG2020} \\
NGC\,5189 & [WO1] & 3.60 & $-$7.34  & 157 & 0.077 & 2000 & 26.20 & 1.09 & \citet{Todt2015b} \\
PB\,6     & [WO1] & 3.56 & $-$7.25  & 157 & 0.082 & 2000 & 26.31 & 1.48 & \citet{Todt2015b} \\ 
PC\,22    & [WO1] & 3.78 & $-$7.36  & 130 & 0.151 & 4500 & 26.66 & 1.49 & \citet{Sabin2022} \\
NGC\,2867 & [WO2] & 3.68 & $-$7.20  & 157 & 0.094 & 2000 & 26.38 & 1.26 & \citet{Todt2015b} \\
NGC\,6905 & [WO2] & 3.90 & $-$6.96  & 139 & 0.153 & 2000 & 26.73 & 1.35 & \citet{GG2022} \\
Hen\,2-55 & [WO3] & 3.43 & $-$7.54  & 126 & 0.101 & 2000 & 26.06 & 1.02 & \citet{Todt2015b} \\
NGC\,6369 & [WO3] & 3.78 & $-$6.78  & 124 & 0.169 & 1600 & 26.83 & 2.10 & \citet{Todt2015b} \\
$[$S71d$]$& [WO3] & 3.91 & $-$6.65  & 166 & 0.113 & 2400 & 27.02 & 3.15 & \citet{Todt2015b} \\
NGC\,1501 & [WO4] & 4.10 & $-$6.80  & 112 & 0.292 & 2000 & 27.03 & 1.20 & \citet{Rubio2022} \\  
\hline
A\,30     & [WC]-PG1159 & 3.90 & $-$7.58 & 115 & 0.226 & 4000 & 26.50 & 0.63 & \citet{Guerrero2012} \\
A\,78     & [WC]-PG1159 & 3.93 & $-$7.70 & 117 & 0.226 & 3100 & 26.27 & 0.35 & \citet{Toala2015} \\
\hline
RaMul\,2  & [WC4-5] & 3.50 & $-6.38$  & 90 & 1.569 & 1000 & 27.51 & 6.52 & \citet{Werner2024} \\ 
NGC\,40   & [WC8]   & 3.85 & $-$6.10  & 64 & 0.681 & 1000 & 28.00 & 5.34 & \citet{Toala2019} \\
SwSt\,1   & [WC9]   & 3.86 & $-$6.25  & 40 & 1.630 & 900  & 27.61 & 3.32 & \citet{Hajduk2020} \\ 
\hline
\end{tabular}
\end{center}
\end{table*}

There have been dedicated studies to characterise the atmosphere and wind properties of WR stars in our Galaxy and the Magellanic Clouds \citep[e.g.,][]{Hamann2006,Sander2012,Hainich2014,Shenar2016,Shenar2019,Tramper2015} and with the advent of {\it Gaia}, reliable distances, and consequently, the determination of these parameters has been improved. \citet{Sander2019} and \citet{Hamann2019} updated their estimates of distant-dependent quantities, like stellar radius ($R_\star$), luminosity ($L$) and mass-loss rates ($\dot{M}$), for single WC, WO and WN stars based on previous analyses with the Potsdam WR (PoWR) NLTE stellar atmosphere code \citep{Grafener2002,Hamann2004,Sander2015}. \citet{Sander2019} used their predictions to establish an $\dot{M}$--$L$ relationship and studied the behaviour of the modified wind momentum \citep[$D_\mathrm{mom}= \dot{M} \varv_\infty \sqrt{R_\star/\mathrm{R}_\odot}$, with $\varv_\infty$ as the stellar wind velocity;][]{Kudritzki1999} for 43 WC stars. In these diagrams, WC stars seem to follow a single sequence regardless of their spectral sub-type. However, this is not the case for the 55 Galactic WN stars analysed by \citet{Hamann2006,Hamann2019} where different correlations appear for early and late WN stars. Unfortunately, WO stars are not numerous to produce similar analyses. However, recent efforts to model these stars have also been presented recently in the literature \citep[e.g.,][]{Aadland2022}

In order to push further our understanding of the WR phenomenon, we present one of the first attempts to study the winds of [WR] stars as part of the WR sample \citep[see][for an overview in the pre-\textit{Gaia} era]{Hamann2010}. [WR] star are considered to be the descendants of low-mass, Sun-like stars. Usually they are surrounded by planetary nebulae, thus they belong to the class of the central stars of planetary nebulae (CSPNe). They are classified using the same methods as for massive WR stars, but their spectral sub-type is written with square brackets to denote their origins as low-mass stars \citep[see, e.g.,][]{Acker2003}.

\citet{Weidmann2020} compiled the most recent census of H-deficient CSPNe and listed a total of 125 [WR]-type CSPNe: 68 [WC], 37 [WO], 8 [WN], 10 [WR] and 2 [WC]-PG1159. Adding such stars to the analysis of the wind properties of WR-type stars helps us to explore the lower luminosity range of $\log_{10}(L$/L$_\odot)< 4.5$, which is otherwise not covered by the Population I WR stars. Besides, it gives us the opportunity to assess the properties of WO-type stars given that [WO]-type stars are more numerous than their massive counterparts. 

In this paper, we use the [WR]-type CSPNe that have been analysed by our group using the PoWR code. We have compiled a sample that includes [WO], [WC] and [WC]-PG1159 stars which will be analysed in conjunction with the massive, presumably single WC and WO stars listed in the literature. The parameters of the sources taken from the literature were corrected using the updated distances based on {\it Gaia} data (see Table~\ref{tab:stars}).

This paper is organised as follows: In Section~\ref{sec:methods} we present our methodology and the sample of [WR]-type stars used here. Our results are presented in Section~\ref{sec:results}. A discussion of our results including comparisons with massive WR stars is presented in Section~\ref{sec:discussion}. Finally, a summary of our results is presented in Section~\ref{sec:summary}.

\section{Methods}
\label{sec:methods}

In the recent years, our group has presented multi-wavelength characterisations of planetary nebulae hosting H-deficient [WR]-type stars. These works included the study of the stellar atmospheres of their central stars using the PoWR code. 

Our group and collaborators have performed stellar atmosphere studies for [WR]-type stars of the [WO], [WC], [WN] and [WC]-PG1159 sequences. Among these are the [WO]-type CSPNe of NGC\,1501, NGC\,2371, NGC\,6369, NGC\,6905 and PC\,22 \citep[][Toal\'{a} et al. in prep.]{GG2020,GG2022,Rubio2022,Sabin2022}, the [WC]-type stars of NGC\,40 \citep{Toala2019}, SwSt\,1 \citep{Hajduk2020}, and the [WC]-PG1159 CSPNe A\,30 and A\,78 \citep{Guerrero2012,Toala2015}. We have also published preliminary results for Hen\,2-55 ([WO3]), NGC\,2867 ([WO2]), NGC\,5189 ([WO1]), PB\,6 ([WO1]) and [S71d] ([WO3]) \citep[see][]{Todt2015b}. 
For all these objects, we recalculated $L$ and hence $R_\star$ and $\dot{M}$ with help of the recent distances from \citet{BJ2021} based on the \textit{Gaia} DR3. In addition, we also include RaMul~2 which was analysed recently in \citet{Werner2024}.

\begin{figure*}
\begin{center} 
\includegraphics[angle=0,width=0.5\linewidth]{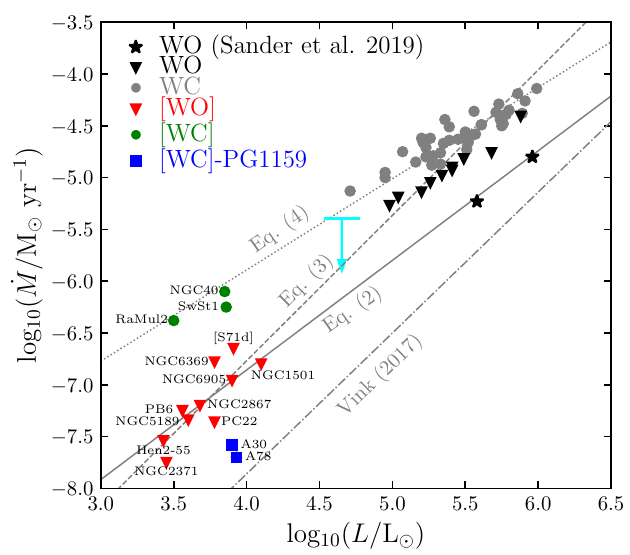}~
\includegraphics[angle=0,width=0.5\linewidth]{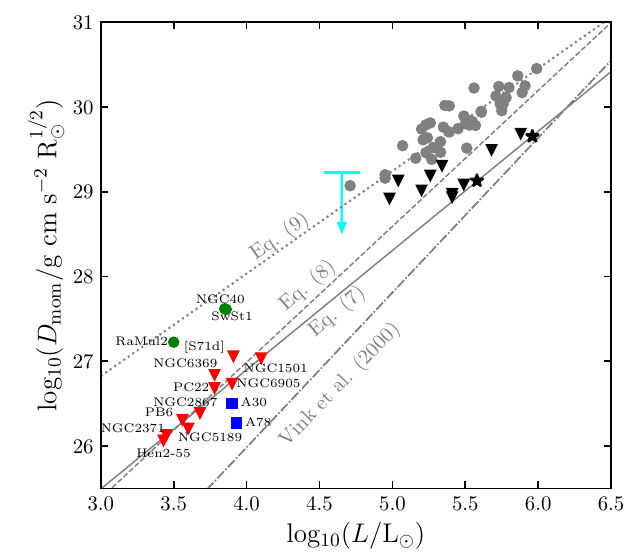}
\caption{Mass-loss rate ($\dot{M}$ - left) and modified wind momentum ($D_\mathrm{mom}$ - right) versus stellar luminosity ($L$). Bullets, triangles and squares symbols represent different spectral types as illustrated by the colour code. Values from WC stars were taken from \citet{Sander2019} and those of Galactic and extragalactic WO stars were taken from \citet{Tramper2015} and \citet{Aadland2022}. For comparison, the two WO stars analysed in \citet{Sander2019} (WR~102 and WR~142) are shown with black stars. Predictions for the double-degenerate merger of Pa\,30 from \citet{Lykou2023} are presented in cyan. The dotted-dashed lines in the left and right panel shows the predictions for optically-thin stars from \citet{Vink2017} and \citet{Vink2000}, respectively, which are expected to be lower limits for optically thick objects such as WR-type stars.}
\label{fig:wind}
\end{center} 
\end{figure*} 

Once we gathered all of the stellar and wind parameters of the different CSPNe in our sample, we follow the same procedure as that of \citet{Sander2019}. We plot our CSPNe into different diagrams combining luminosity $L$, mass-loss rate $\dot{M}$, stellar temperature $T_\mathrm{eff}$, stellar radius $R_\star$, stellar wind velocity $\varv_\infty$, modified wind momentum $D_\mathrm{mom}$ and the wind efficiency parameter $\eta$ (see Sec.~\ref{sec:other}) to search for dependence patterns. The values of these parameters are also listed in Table~\ref{tab:stars} for all [WR]-type stars used here. We give the stellar temperature $T_\mathrm{eff}$ and the stellar radius $R_\star$ at a radial optical of $\tau_\mathrm{Rosseland}=2/3$ to allow a comparison with the results of \citet{Aadland2022}. All of the fits to the data were obtained by applying linear regression procedures.

We note that the [WC]-PG1159 stars of A\,30 and A\,78 tend to be outliers in the analysis presented in the following sections and, consequently, we did not include them in any of the linear regression fits presented below. Instead, these are only used for comparison and discussion. The [WC]-PG1159 stars mark a spectroscopic transition type between emission- and absorption-line stars in the low-mass regime. The CSPNe of A\,30 and A\,78 are thought to have experienced a so-called very late thermal pulse \citep[VLTP or a born-again event; see][]{Schoenberner1979,Iben1983} when descending the WD cooling track, rendering them [WR]-type CSPN \citep{Todt2015b,Todt2015c}. We note that this scenario does not have a known counterpart in the massive star regime.

\section{Results}
\label{sec:results}

\subsection{$\dot{M}$ versus $L$}
  \label{sec:mdotl}

We start by plotting our sample of [WR] stars in a $\dot{M}$--$L$ diagram in the left panel of Fig.~\ref{fig:wind}. Although not numerous, [WO] stars seem to align and suggest a sequence. After performing a linear regression, we found that the [WO] stars can be fitted by the relationship
\begin{equation}
   \log_{10} \left(\frac{\dot{M}}{\mathrm{M}_\odot~\mathrm{yr}^{-1}} \right) = (1.30\pm0.27) \log_{10} \left(\frac{L}{\mathrm{L}_\odot}\right) - (11.99\pm0.99). 
   \label{eq:WO0}
\end{equation}
Although the limited number of [WC] and [WC]-PG1159 stars do not allow us to perform such a fit, interesting results can be envisaged after including massive WO and WC stars analysed in the literature to Fig.~\ref{fig:wind}. In fact, the low-mass [WO] stars align with the predictions for the two massive WO stars analysed in \citet{Sander2019} (i.e., WR\,102 and WR\,142), producing a single sequence. A linear regression to [WO] and WO (hereinafter [WO]+WO) results in
\begin{equation}
    \log_{10} \left(\frac{\dot{M}}{\mathrm{M}_\odot~\mathrm{yr}^{-1}} \right) = (1.06\pm 0.07) \log_{10} \left(\frac{L}{\mathrm{L}_\odot}\right) - (11.14\pm 0.29).
    \label{eq:WO1}
\end{equation}
When computing a linear regression simultaneously fitting [WO] and WO from other works, such as \citet{Tramper2015} and \citet{Aadland2022}, we obtain instead
\begin{equation}
    \log_{10} \left(\frac{\dot{M}}{\mathrm{M}_\odot~\mathrm{yr}^{-1}} \right) = (1.40\pm0.31) \log_{10} \left(\frac{L}{\mathrm{L}_\odot}\right) - (12.38\pm1.16),
    \label{eq:WO}
\end{equation}
\noindent which is, within errorbars, more similar to that fit obtained for [WO] stars only in Eq.~(\ref{eq:WO0}). We attribute this difference to the different model approaches with the WO models from \citet{Sander2012,Sander2019}, probably overestimating the amount of the oxygen for the targets. Eqs.~(\ref{eq:WO1}) and (\ref{eq:WO}) are overplotted in the left panel of Fig.~\ref{fig:wind} with solid and dashed lines, respectively.

The situation with WC and [WC] (hereinafter [WC]+WC) stars is inverted with regards to the sample sizes. There are many massive WC stars but just a few [WC] in our sample. If a similar linear regression analysis is performed to [WC]+WC stars, we obtain
\begin{equation}
    \log_{10} \left(\frac{\dot{M}}{\mathrm{M}_\odot~\mathrm{yr}^{-1}} \right) = (0.88\pm0.03) \log_{10} \left(\frac{L}{\mathrm{L}_\odot}\right) - (9.42\pm0.17).
    \label{eq:WC}
\end{equation}
\noindent Eq.~(\ref{eq:WC}) is plotted with a dotted line in the left panel of Fig.~\ref{fig:wind}. For comparison, a linear regression fit only to massive WC stars results in 
\begin{equation}
    \log_{10} \left(\frac{\dot{M}}{\mathrm{M}_\odot~\mathrm{yr}^{-1}} \right) = (0.69\pm0.05) \log_{10} \left(\frac{L}{\mathrm{L}_\odot}\right) - (8.38\pm0.28).
    \label{eq:WC_only}
\end{equation}
\noindent Eq.~(\ref{eq:WC_only}) is almost exactly the same as equation~(7) of \citet{Sander2019}, which employed the earlier \textit{Gaia} DR2 distances, but notably different than the combined Eq.~(\ref{eq:WC}).

\subsection{${D}_\mathrm{mom}$ versus $L$}

We further explored the $D_\mathrm{mom}$--$L$ diagram in the right panel of Fig.~\ref{fig:wind}. Similarly to the $\dot{M}$--$L$ panel, two independent sequences can be identified between [WO]+WO and [WC]+WC types, where those of the O sequence fall below those of the C one.

\begin{figure*}
\begin{center} 
\includegraphics[angle=0,width=0.5\linewidth]{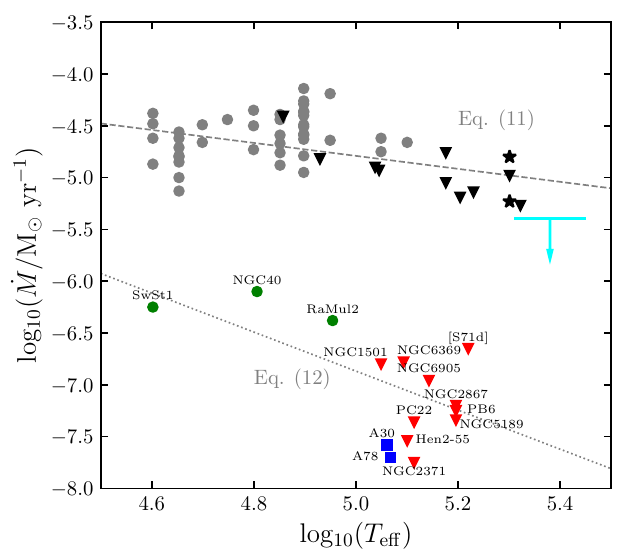}~
\includegraphics[angle=0,width=0.5\linewidth]{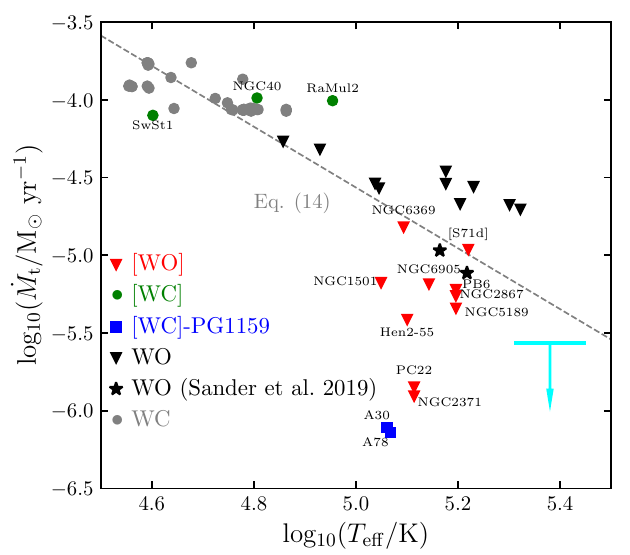}\\
\caption{{\bf Left:} Effective temperature $T_\mathrm{eff}$ versus mass-loss rate $\dot{M}$. Two different sequences can be envisaged, one for massive WR stars (Eq.~\ref{eq:massloss1}) and another for low-mass [WR] stars (Eq.~\ref{eq:massloss2}). {\bf Right:} Effective temperature $T_\mathrm{eff}$ versus transformed mass-loss rate $\dot{M}_\mathrm{t}$. A linear regression fit to all stars (Eq.~\ref{eq:trans_massive}) is shown with a dashed line. Predictions for the double-degenerate merger of Pa\,30 from \citet{Lykou2023} are presented as upper limit (cyan) arrows in both panels. The symbols have the same meaning as in Fig.~\ref{fig:wind}.}
\label{fig:trans_mass}
\end{center} 
\end{figure*}

A linear regression to only the [WO] stars results in 
\begin{equation}
    \log_{10} \left(\frac{D_\mathrm{mom}}{\mathrm{g}~\mathrm{cm}~\mathrm{s}^{-2}~\mathrm{R_\odot}^{1/2}}\right) = (1.61\pm 0.19) \log_{10} \left(\frac{L}{\mathrm{L}_\odot}\right) + (20.56 \pm 0.71),
    \label{eq:WO_mom0}
\end{equation}
\noindent while a fit to [WO]+WO stars only taking into account WR\,102 and WR\,142 from \citet{Sander2019} results in 
\begin{equation}
    \log_{10} \left(\frac{D_\mathrm{mom}}{\mathrm{g}~\mathrm{cm}~\mathrm{s}^{-2}~\mathrm{R_\odot}^{1/2}}\right) = (1.40\pm 0.05) \log_{10} \left(\frac{L}{\mathrm{L}_\odot}\right) + (21.27 \pm 0.22),
    \label{eq:WO_mom1}   
\end{equation}
\noindent but taking into account WO from other works yields
\begin{equation}
    \log_{10} \left(\frac{D_\mathrm{mom}}{\mathrm{g}~\mathrm{cm}~\mathrm{s}^{-2}~\mathrm{R_\odot}^{1/2}}\right) = (1.56\pm0.07) \log_{10} \left(\frac{L}{\mathrm{L}_\odot}\right) + (20.75\pm0.31).
    \label{eq:WO_mom}   
\end{equation}
\noindent Eqs.~(\ref{eq:WO_mom1}) and (\ref{eq:WO_mom}) are illustrated with solid and dashed lines in the right panel of Fig.~\ref{fig:wind}, respectively. 

A similar regression to the [WC]+WC stars results in 
\begin{equation}
    \log_{10} \left(\frac{D_\mathrm{mom}}{\mathrm{g}~\mathrm{cm}~\mathrm{s}^{-2}~\mathrm{R_\odot}^{1/2}}\right) = (1.21\pm0.16) \log_{10} \left(\frac{L}{\mathrm{L}_\odot}\right) + (23.19\pm0.25),
    \label{eq:WC_mom}
\end{equation}
\noindent which is plotted in the right panel of Fig.~\ref{fig:wind} as a dotted line. For comparison, a linear regression performed only on WC stars results in
\begin{equation}
    \log_{10} \left(\frac{D_\mathrm{mom}}{\mathrm{g}~\mathrm{cm}~\mathrm{s}^{-2}~\mathrm{R_\odot}^{1/2}}\right) = (1.02\pm0.08) \log_{10} \left(\frac{L}{\mathrm{L}_\odot}\right) + (24.22\pm0.43).
    \label{eq:WC_mom_only}
\end{equation}

\subsection{$\dot{M}$ and $\varv_\infty$ versus $T_\mathrm{eff}$}

The other independent parameter is the effective temperature $T_\mathrm{eff}$. Unveiling its relationship with the mass-loss rate $\dot{M}$ is of highest interest as it could help eventually obtain realistic temperature estimates for evolved stars with dense winds, particularly, WR-type stars. 

Fig.~\ref{fig:trans_mass} (left panel) illustrates that, in general, massive WR stars exhibit larger mass-loss rate values than their [WR] counterparts with the same $T_\mathrm{eff}$ (or spectral type). Plotting the sample of [WR] stars analysed here in the $\dot{M}$--$T_\mathrm{eff}$ diagram shows that 
there are evidently two independent sequences defined by massive and low-mass stars. Linear regression fits to these sequences resulted in
\begin{equation}
    \log_{10} \left(\frac{\dot{M}}{\mathrm{M}_\odot~\mathrm{yr}^{-1}} \right) = (-0.63\pm0.15) \log_{10} \left(\frac{T_\mathrm{eff}}{\mathrm{K}}\right) + (1.66\pm0.78),
    \label{eq:massloss1}
\end{equation}
\noindent for massive WR stars and 
\begin{equation}
    \log_{10} \left(\frac{\dot{M}}{\mathrm{M}_\odot~\mathrm{yr}^{-1}} \right) = (-1.88\pm0.63) \log_{10} \left(\frac{T_\mathrm{eff}}{\mathrm{K}}\right) + (2.55\pm3.00),
    \label{eq:massloss2}
\end{equation}
\noindent for low-mass [WR] stars. Eq.~(\ref{eq:massloss1}) and (\ref{eq:massloss2}) are plotted alongside the data in the left panel of Fig.~\ref{fig:trans_mass} with dashed and dotted lines, respectively.

Another quantity that is worth analysing is the transformed mass-loss rate $\dot{M}_\mathrm{t}$ defined by \citet[][]{Grafener2013} as: 
\begin{equation}
\dot{M}_\mathrm{t} = \dot{M} \sqrt{D} \left(\frac{1000~\mathrm{km}~\mathrm{s}^{-1}}{\varv_\infty}\right)\left( \frac{10^{6}~\mathrm{L}_\odot}{L}\right)^{3/4},
\end{equation}
\noindent which describes the mass-loss rate a star would have if it had an unclumped ($D = 1$) wind with a terminal velocity of $1000\,\mathrm{km}\,\mathrm{s}^{-1}$ and a luminosity of $10^{6}\,\mathrm{L}_\odot$. Thereby, this parameter eliminates the explicit dependence on the observed $L$-regimes.

\begin{figure}
\begin{center} 
\includegraphics[angle=0,width=\linewidth]{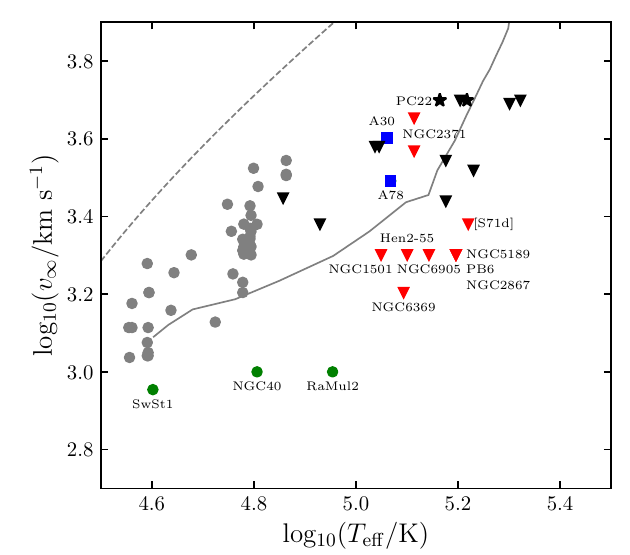}
\caption{$\varv_\infty$--$T_\mathrm{eff}$ diagram. The grey solid line corresponds to the $\varv_{\infty}$--$T_\mathrm{eff}$ relationship from \citet{Sander2023} that illustrates the case of a 20~M$_\odot$ WC star. For comparison the dashed line shows the results for a fit to OB stars in our Galaxy presented in \citet{Hawcroft2023}. The symbols have the same meaning as that of Fig.~\ref{fig:wind}.}
\label{fig:velocity}
\end{center} 
\end{figure}

\begin{figure}
\begin{center} 
\includegraphics[angle=0,width=\linewidth]{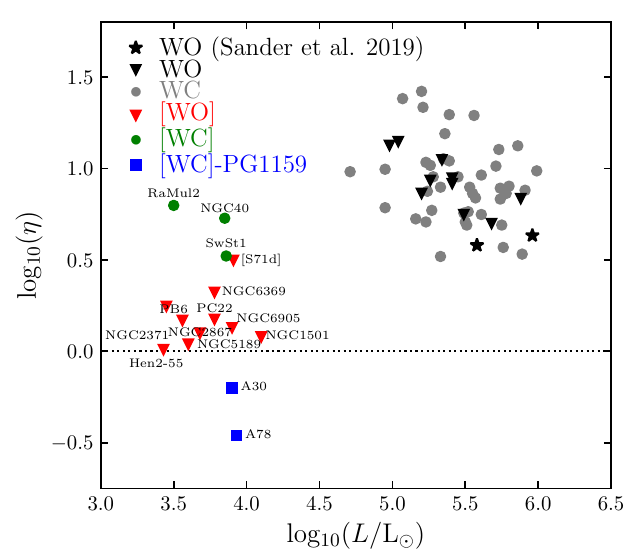}\\
\includegraphics[angle=0,width=\linewidth]{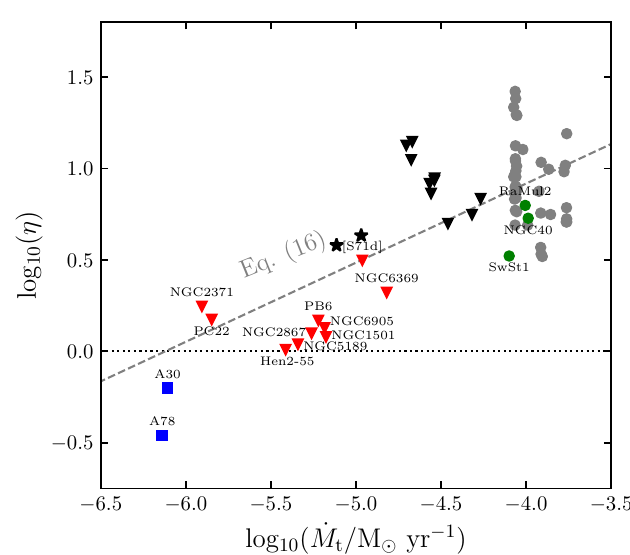}
\caption{{\bf Top:} $\eta$--$L$ diagram. {\bf Bottom:} $\eta$--$\dot{M}_\mathrm{t}$ diagram (see Eq.~\ref{eq:mt}). The colour and symbol codes are the same as for Fig.~\ref{fig:wind}. The horizontal dotted line in both panels corresponds to $\eta=1$.}
\label{fig:eta}
\end{center} 
\end{figure}

The right panel of Fig.\,\ref{fig:trans_mass} plots all WR-type stars and, unlike the left panel, the $\dot{M}_\mathrm{t}$--$T_\mathrm{eff}$ diagram suggests a single sequence that includes massive and low-mass WR-type stars.  The [WC] and WC clearly populate the same parameter space in this plane. The [WO] and WO stars are also broadly located in a the same region of the diagram, but there seems to be a tendency towards slightly lower values of $\dot{M}_\mathrm{t}$ for the [WO] stars. Moreover, the [WC]-PG1159 stars (A\,30 and A\,78) are clearly below the trend, with two further [WO1] objects (PC\,22 and NGC\,2371) also deviating considerably from the linear regression fit
\begin{equation}
    \log_{10} \left(\frac{\dot{M}_\mathrm{t}}{\mathrm{M}_\odot~\mathrm{yr}^{-1}} \right) = (-1.95\pm0.16) \log_{10} \left(\frac{T_\mathrm{eff}}{\mathrm{K}}\right) + (5.22\pm0.80),
    \label{eq:trans_massive}
\end{equation}
\noindent which is illustrated in the right panel of Fig.~\ref{fig:trans_mass} as a dashed grey line. We further plot the WO-like object Pa~30 with parameters from \citet{Lykou2023}, which also seems to fall a bit below the trend, despite its good alignments in the $\dot{M}$-$L$ and $D_\text{mom}$-$L$ planes. Yet, there is a considerable uncertainty in the results of Pa 30 and also several of the low- and high-mass WO stars, which could potentially overemphasise the scatter in the WO domain. For example, just a slightly stepper trend could lead to an alignment of Pa 30 with the overall trend. A homogeneous analysis of low- and high-mass WO stars will be necessary to get a better handle on $\dot{M}_\mathrm{t}$ versus $T_\mathrm{eff}$, which we will discuss further from a theoretical point of view in Sect.\,\ref{sec:discussion}.

Further exploration of $T_\mathrm{eff}$ with respect to the stellar wind velocity $\varv_\infty$ is illustrated in Fig.~\ref{fig:velocity}. This panel shows a scattered landscape with massive WR stars exhibiting typically larger stellar wind velocities than [WR] stars with the same $T_\mathrm{eff}$.

\subsection{Wind efficiency parameter} 
\label{sec:other}

We also computed the wind efficiency parameter $\eta$ for all stars studied here, which is defined as
\begin{equation}
    \eta = \frac{\dot{M} \varv_\infty}{L/c}.
    \label{eq:eta}
\end{equation}
\noindent Different plots comparing $\eta$ with other parameters were inspected, but we only presented two of them in Fig.~\ref{fig:eta}. The top panel illustrates that massive stars have usually higher wind efficiencies, even when accounting for the difference in luminosity. A notable exception are the [WC] stars in our sample (see Fig.~\ref{fig:eta}) which have very similar $\eta$-values to those of their massive counterparts. Interestingly, the CSPNe for which the born-again origin is most probable (A\,30 and A\,78) are the only objects in our sample with $\eta < 1$ (see Fig.~\ref{fig:eta} and Table~\ref{tab:stars}).

The $\eta$--$L$ panel in Fig.~\ref{fig:eta} (top) further illustrates the luminosity ranges between [WR] and WR stars, revealing that the stars in this panel fall into two different regions of the diagram. 
When comparing $\eta$ with other parameters, we see that this is merely a consequence of the missing detections in a certain luminosity regime around $\log_{10}(L/\mathrm{L}_\odot) \approx 4.5$.

In the bottom panel of Fig.~\ref{fig:eta}, we show $\eta$ as a function of $\dot{M}_\mathrm{t}$. Even though the number of [WR] stars in our sample is not large, the bottom panel of Fig.\,\ref{fig:eta} suggests a single sequence in the $\eta$--$\dot{M}_\mathrm{t}$ plane which seems to smoothly connect WR-type stars of the O sequence in the low $\dot{M}_\mathrm{t}$ regime with those of the C sequence towards large $\dot{M}_\mathrm{t}$ values. Similar to the $\dot{M}_\mathrm{t}$--$T_\mathrm{eff}$ diagram of Fig.~\ref{fig:trans_mass}, [WC] stars are located in the same region of the $\eta$-$\dot{M}_\mathrm{t}$ diagram with WC stars, and the same situation is shown by stars of the O-sequence (WO and [WO] stars). A linear regression to the stars in the $\eta$--$\dot{M}_\mathrm{t}$ diagram results in
\begin{equation}
\log_{10}(\eta) = (0.43\pm0.06)\log_{10}(\dot{M}_\mathrm{t}) + (2.65\pm0.25).
\label{eq:mt}
\end{equation}

\section{Discussion}
\label{sec:discussion}

Including low-mass [WR] stars to the predictions of the atmosphere and wind properties of WR-type stars enables us to sample the low luminosity WR regime. Despite an absence of objects around $\log_{10}(L/\mathrm{L}_\odot) \approx 4.5$, the low-mass [WC] stars follow a similar trend as the massive WC stars in the diagrams presented in Fig.~\ref{fig:wind}.

By including [WC] stars in the linear regression calculations for $\dot{M}$--$L$ we obtain a dependence of $\dot{M} \propto L^{0.88\pm0.03}$ (see Eq.~\ref{eq:WC}), which gives a slightly steeper correlation than that one by \citet{Sander2019} of $\dot{M} \propto L^{0.68}$ (see also Eq.~\ref{eq:WC_only}) only accounting for massive WC stars. Interestingly, our new result for [WC]+WC stars is very similar to that from \citet{NugisLamers2000} of $\dot{M} \propto L^{0.84}$ inferred for WC stars. 
\citet{Hamann2010} discussed the $\dot{M}$--$L$ relation for hot massive and low mass stars, including also [WC] and [WO] stars. Yet, \citet{Hamann2010} had to adopt a fixed luminosity of $6000\,\text{L}_\odot$ for most of the [WC] and [WO] stars as this was in the pre-\textit{Gaia} era where most objects did not have a reliable distance estimate. Therefore, only a trend for a sample of extreme helium stars and subdwarfs could be inferred, yielding $\dot{M} \propto L^{1.5}$, which is much steeper than our new relations.

In order to solidify our obtained results, more [WC] stars will need to be analysed, but the fact that these low-mass stars align with the trends from massive WC stars indicates that including low-mass objects can help us explore a wider luminosity range of C-rich WR-type stars, forming a single sequence for these type of objects, independent of the mass regime.

Our analysis of [WC]+WC stars suggest a dependency of $D_\mathrm{mom} \propto L^{1.21\pm0.16}$, which is within error bars the same as the found for massive WC stars ($D_\mathrm{mom} \propto L^{1.02\pm0.08}$; see Eq.~\ref{eq:WC_mom_only}). We note that this relationship is not exactly the same as that one reported by \citet{Sander2019}, as all $D_\mathrm{mom}$ values in \citet{Sander2019} are 0.5 dex higher than those computed using the raw definition of $D_\mathrm{mom}$(=$\dot{M} \varv_\infty \sqrt{R_\star/\mathrm{R}_\odot}$). It appears that this discrepancy is due to an explicit inclusion of the clumping parameter in their $D_\mathrm{mom}$ figures.
We conclude that when including [WC] stars in the $D_\mathrm{mom}$-calculations, the presumed exponent $\alpha_\text{eff}$ from $D_\text{mom} \propto L^{1/\alpha_\text{eff}}$ \citep[see, e.g.,][]{Kudritzki1999} slightly decreases from $\alpha_\text{eff} \approx 1.0$ \citep[cf.\ Sect.\,3.3 in][]{Sander2019} to $\alpha_\text{eff} \approx 0.83$. Given the scatter, this difference is propably not significant and in any case $\alpha_\text{eff}$ would be larger than determined for stars with optically thin winds in \citet{Kudritzki1999}.

The inclusion of low-mass [WO] stars enabled us to get a larger sample of WO-type stars for the first time as the [WO] stars are numerous enough to perform a linear regression on their properties as a sample. For the low-mass regime alone, we found dependencies of $\dot{M} \propto L^{1.30\pm0.27}$ (Eq.~\ref{eq:WO0}) and $D_\mathrm{mom} \propto L^{1.61\pm0.19}$ (Eq.~\ref{eq:WO_mom0}) for the mass-loss rate and modified wind momentum, respectively. 
Adding the massive WO stars with parameters reported by \citet{Tramper2015} and \citet{Aadland2022}, we obtain $\dot{M} \propto L^{1.40\pm0.31}$ (Eq.~\ref{eq:WO}) and $D_\mathrm{mom} \propto L^{1.56\pm0.07}$ (Eq.~\ref{eq:WO_mom}) for the combined [WO]+WO sample. This is essentially within the error bars of the low-mass results and we thus conclude that their winds adhere to similar physical scalings.

Yet, there is a difference in $L$ and $\dot{M}$ between the work of \citet{Sander2019} and that of \citet{Tramper2015} and \citet{Aadland2022} which cause small discrepancies when deriving predictions based on either of the samples. So far, a coherent analysis of the massive WO stars beyond the LMC sample by \citet{Aadland2022} is missing. On the one hand, \citet{Tramper2015} analysed more WO stars than \citet{Sander2019}, but their work does not account for Fe ions higher than Fe\,{\sc x} which can significantly affect the estimation of $T_\mathrm{eff}$ and possibly is one of the roots of the discrepancies. On the other hand, the models used in \citet{Sander2019} are more rudimentary as they are from an earlier generation of models. It will thus be important to eventually reanalyse the WO stars in \citet{Tramper2015} by means of current atmosphere models to assess the differences between the methods and obtain a more coherent picture. 

Beside remaining caveats, we could demonstrate in this work that [WO]+WO and [WC]+WC stars represent two different regimes in the $\dot{M}$--$L$ and $D_\mathrm{mom}$--$L$ diagrams.
Their winds are stronger than those predicted for massive OB stars and envelope-stripped OB stars in \citet{Vink2000} and \citet{Vink2017}. These relationships are also plotted in Fig.~\ref{fig:wind} with dash-dotted lines.

We showed that massive and low-mass stars of the WR-type aligned in two different sequences defined by the $\dot{M}_\mathrm{t}$--$T_\mathrm{eff}$ (Eq.~\ref{eq:trans_massive}) and $\eta$--$\dot{M}_\mathrm{t}$ (Eq.~\ref{eq:mt}) diagrams. This suggest that the transformed mass-loss rate $\dot{M}_\mathrm{t}$ might be the best way to compare the two apparently disconnected populations of massive WR and low-mass [WR] stars. The exceptions in the $\dot{M}_\mathrm{t}$--$T_\mathrm{eff}$ diagram seem to be the [WO1] stars PC\,22 and NGC\,2371, in addition to the [WC]-PG1159 objects. But we note that this might be suggestive of a more fundamental property of the WR phenomenon where more dense winds are inherently associated to higher wind efficiencies which is also supported by theoretical calculations for massive-star WR winds \citep[][]{Sander2020b,Sander2023}.

Given that definitions of $\eta$ and $\dot{M}_\mathrm{t}$ use the same variables, except of the clumping factor $D$, the observed trend in the bottom panel of Fig.~\ref{fig:eta} might just be a consequence of this. The inclusion of WN and [WN] stars in a future analysis will help us peer deeper into this. We note that the $\eta$--$L$ panel (Fig.~\ref{fig:eta} - top) also suggest a similar trend, but the scatter is larger than that of the $\eta$--$\dot{M}_\mathrm{t}$ diagram (Fig.~\ref{fig:eta} - bottom).

The prediction of $T_\text{eff}$ for optically thick winds marks a particular challenge for stellar evolution models as the calculation of a detailed atmosphere model is usually not feasible within the framework of a stellar evolution calculation. For example, \citet{Groh2014} calculated CMFGEN models for a single 60\,M$_\odot$ track, demonstrating the difference between simpler estimates and a proper connection to a detailed atmosphere calculation \citep[see also][]{Schaerer1996,HegerLanger1996}. A systematic connection between a quantity like $\dot{M}_\text{t}$ which is set at the base of the wind and  $T_\text{eff}$, which is located in the expanding layers could potentially simplify this task, but more parameter studies are required to see if this is actually possible.

When comparing the empirically obtained values, we found that all WR-type stars form a single sequence in the $\dot{M}_\mathrm{t}$--$T_\mathrm{eff}$ diagram of Fig.~\ref{fig:trans_mass} with a dependence of $\dot{M}_\mathrm{t} \propto T_\mathrm{eff}^{-1.95\pm0.16}$ (see Eq.~\ref{eq:trans_massive}). For comparison, we now consider theoretical studies on the $\dot{M}_\mathrm{t}$--$T_\mathrm{eff}$ dependency. The model sequences from \citet{Sander2020b} yield
\begin{equation}
  \log_{10} \left(\frac{\dot{M}_\mathrm{t}}{\mathrm{M}_\odot~\mathrm{yr}^{-1}} \right) = (-1.5\pm0.02) \log_{10} \left(\frac{T_\mathrm{eff}}{\text{K}}\right) + (3.32\pm0.09),
\end{equation}
\noindent while the recent sequences in \citet{Sander2023} result in
\begin{equation}
  \log_{10} \left(\frac{\dot{M}_\mathrm{t}}{\mathrm{M}_\odot~\mathrm{yr}^{-1}} \right) = (-2.04\pm0.04) \log_{10} \left(\frac{T_\mathrm{eff}}{\text{K}}\right) + (5.94\pm0.16).
  \label{eq:trans_mass_sander}
\end{equation}
\noindent The slope of Eq.~(\ref{eq:trans_mass_sander}) is almost identical to our obtained empirical trend in Eq.~(\ref{eq:trans_massive}). In both cases $\dot{M}_\mathrm{t} \propto T_\mathrm{eff}^{-2}$, hinting at a trend that might be universal for stars exhibiting the WR phenomenon.

The scaling of $\varv_\infty$ with $T_\text{eff}$ is a well established phenomenon in the framework of OB stars \citep[e.g.,][]{Evans2004,Garcia2014,Hawcroft2023}. It is theoretically expected due to the scaling of $\varv_\infty$ with the escape velocity  \citep{Castor1975,Pauldrach1986}, but due to the uncertainties in determining the stellar masses, the direct correlation of $\varv_{\infty}(T_\text{eff})$ is usually more robust \citep{VinkSander2021}. The relation is not strictly followed as outliers can be produced by stars with higher $\dot{M}$ which usually show lower values of $\varv_\infty$ than expected from the relation. The empirical collection of data also hints at a metallicity dependence in $\varv_{\infty}(T_\text{eff})$ due a $Z$-dependence of $\varv_\infty$, but the derived slopes depend a lot of the employed sample \citep[e.g.,][]{Hawcroft2023}. For WR stars with optically thick winds, theoretical calculations by \citet{Sander2023} predict a shallower slope of $\varv_{\infty}(T_\text{eff})$ compared to the regime of OB stars. While massive WR stars exhibit typically larger stellar wind velocities and the slope is shallower than for OB stars, there seems to be a separation of the more massive WR stars showing higher values of $\varv_\infty$ for the same $T_\text{eff}$ than the bulk of the low-mass sample.

Fig.~\ref{fig:velocity} shows the predicted terminal wind velocity for a hydrodynamically-consistent stellar atmosphere model of a WC star with 20~M$_\odot$, calculated in  \citet{Sander2023} (solid gray line). The shown model curve has a notable kink at $\log_{10}(T_\mathrm{eff}/\mathrm{K})\approx5.1$, that marks the transition from a shallower slope in the optically thick wind regime to a steeper slope for optically thin winds. Depending on the adopted mass and metallicity, \citet{Sander2023} showed that this transition varies from $\log_{10}(T_\mathrm{eff}/\mathrm{K})\approx5.05$ to 5.15.
While our sample is limited, most of our [WR] stars are located below the predicted $\varv_\infty$ for massive WC stars. With [WC] stars located in the region of the diagram where optically thick winds are expected, and [WO] stars in the region of optically thin winds -- together with their massive WO counterparts. However, there are some notable exceptions of [WR]-type stars being located in the same region as the WO-type stars with their optically thin winds. Particularly, the [WC]-PG1159 stars A\,30 and A\,78, which have severely weaker winds as also reflected by very low values of $\dot{M}_\text{t}$ and $\eta$, placing them in the optically-thin regime. Moreover, the [WO]-type stars  NGC\,2371 and PC\,22 turn out to be located in the same region as A\,30 and A\,78 in Fig.~\ref{fig:trans_mass} with regards to not only $\varv_\infty$, but also $\dot{M}_\text{t}$, albeit their wind efficiencies $\eta$ are notably higher than those of the [WC]-PG1159 stars. In order to understand this, hydrodynamically-consistent atmosphere models of [WC]- and [WO]-stars would be required, which is beyond the scope of the present study.

Three of our [WO] stars -- NGC\,2371, NGC5189 and Hen\,2-55 -- have values of $\eta \approx 1$, placing them close to the single scattering limit, which is commonly associated with the transition from optically thin to optically thick winds, i.e., stellar winds where the sonic point is located close below $\tau_\mathrm{F} = 1$, where $\tau_\mathrm{F}$ is the total flux-weighted mean optical depth. Stars in this transition regime enable us to compare our empirical insights with the concept of the transition mass-loss rate $\dot{M}_\text{trans}$ (not to be confused with the transformed mass-loss rate $\dot{M}_\text{t}$), defined as
\begin{equation}
  \label{eq:mdtrans}
  \dot{M}_\text{trans} = \frac{L/c}{\varv_\infty}   
\end{equation}
\noindent and introduced by \citet{Vink2012}. 
Based on a simple integration of the 1D stationary hydrodynamic equation of motion with few simplifications \citep[see][for details]{Vink2012}, Eq.\,\eqref{eq:mdtrans} provides a direct way to calculate the mass-loss rate from two quantities that can be determined empirically and do not depend on uncertain assumptions such as the distribution of inhomogeneities in the wind (clumping).
This expression is valid in the regime with $\eta = f \cdot \tau_\mathrm{F}(R_\text{sonic}) = 1$ where $f$ is on the order of unity with typical values around $f \approx 0.5$ to $0.6$ when examining numerical models of massive stars \citep[see, e.g.,][and references therein]{Sabhahit2023}. This implies that Eq.\,\eqref{eq:mdtrans} would be exact at $\eta \approx 0.6$. Whether this holds also for the very different mass regime of low-mass [WR]-stars has never been tested so far. Thus, in Table\,\ref{tab:mdtrans} we provide the transition mass-loss rate values $\dot{M}_\mathrm{trans}$ predicted by Eq.\,\eqref{eq:mdtrans} and compare them to the values derived by quantitative spectral analysis listed in Table~\ref{tab:stars}.

\begin{table}
\begin{center}
\caption{Estimated transition mass-loss rates $\dot{M}_\mathrm{trans}$ for [WR] and [WC]-PG1159 stars close to the single-scattering limit ($\eta \lesssim 1$).}
\setlength{\tabcolsep}{0.9\tabcolsep}  
\label{tab:mdtrans}
\begin{tabular}{lccccc} 
\hline
Name      & $\eta$ & $\log_{10}(L)$ &  $\varv_\infty$ & $\log_{10}(\dot{M})$  &  $\log_{10}(\dot{M}_\text{trans})$ \\
          &        & (L$_\odot$)& (km~s$^{-1}$)& (M$_\odot$~yr$^{-1}$)& (M$_\odot$~yr$^{-1}$)  \\
\hline
NGC\,2371 & 1.11 & 3.45 & 3700 & $-$7.75  & $-$7.81 \\
NGC\,5189 & 1.09 & 3.60 & 2000 & $-$7.34  & $-$7.40 \\
Hen\,2-55 & 1.02 & 3.43 & 2000 & $-$7.54  & $-$7.57 \\
\hline
A\,30     & 0.63 & 3.90 & 4000 & $-$7.58 &  $-$7.40 \\
A\,78     & 0.35 & 3.93 & 3100 & $-$7.70 &  $-$7.26 \\
\hline
\end{tabular}
\end{center}
\end{table}

Table~\ref{tab:mdtrans} shows that the predictions for the [WC]-PG1159 stars differ actually more from the empirically determined values than those for the [WO] stars with $\eta \gtrsim 1$. While we cannot draw definite conclusions without investigating also the flux-weighted optical depth at the sonic point of these stars, our findings indicate that the factor $f$ between $\tau_\mathrm{F}(R_\text{sonic})$ and $\eta$ is very close to unity for the low-mass [WR] regime. Hence, for low-mass WR stars of same wind efficiency as massive WR stars, a larger wind optical depth is necessary than for their massive counterparts. Yet, the good agreement between theory and empirical analysis in the transition regime indicates that our empirically derived $\dot{M}$-values seem to be reasonably robust and thus also the relations obtained in Sect.\,\ref{sec:mdotl} are well anchored.

Finally, it is also convenient to contrast our findings with the properties of the WO-like central star of the supernova remnant Pa~30 \citep[see][]{Gvaramadze2019,Oskinova2020}. This nebula has been classified as SN\,2002cx-like, very likely the product of a double-degenerate scenario \citep{Kashyap2018}. In a recent paper, \citet{Lykou2023} presented the analysis of this ultra-stripped star using models produced with the CMFGEN code. They found the best-fit stellar and wind parameters to be $\log_{10}(L/\text{L}_{\odot})=4.65^{+0.13}_{-0.17}$, $\log_{10}(T_\mathrm{eff}/\text{K})=5.38^{+0.07}_{-0.08}$, $R_{\star} \leqslant 0.2~\text{R}_\odot$, $\dot{M}\leqslant4\times10^{-6}~\text{M}_\odot~\text{yr}^{-1}$ and $\varv_\infty\approx15,000~\text{km~s}^{-1}$ \citep[see table 1 in][]{Lykou2023}.

The WR-type object in Pa~30 falls in the gap between massive WR and the [WR] stars in the panels of Fig.~\ref{fig:wind}, at $\log_{10}(L/\text{L}_\odot)\approx 4.5$, suggesting that WR-like stars in this luminosity range might only arise from unusual channels represented by this ultra-stripped object. 
Given that the estimates for its $\varv_\infty$ are quite high, its wind efficiency $\eta=90.35$, is much larger than for the other objects of our sample, hence we do not include it in our inferred  relations for the properties of WR-type winds. 
Despite its high value of $\varv_\infty$ and $\eta$, Pa~30 follows more or less the trend of the objects in the $\dot{M}_\mathrm{t}$--$T_\mathrm{eff}$ diagram (see Fig.~\ref{fig:trans_mass} - right panel). While it is too much off the scale to explicitly plot Pa~30 in Fig.~\ref{fig:velocity}, its position in the $\varv_\infty(T_\mathrm{eff})$-diagram actually aligns very well with the extrapolated theoretical curve in the optically thin regime. We therefore conclude that despite its spectrum with very strong emission lines and the huge $\eta$-value -- indicating very efficient multiple scatting, the wind of Pa~30 might not be optically thick over the whole spectral range and is qualitatively similar to those of other WO-type objects, albeit with more extreme properties than usually observed.

\subsection{Fundamental properties of the WR phenomenon}

\begin{figure*}
\begin{center} 
\includegraphics[angle=0,width=0.9\linewidth]{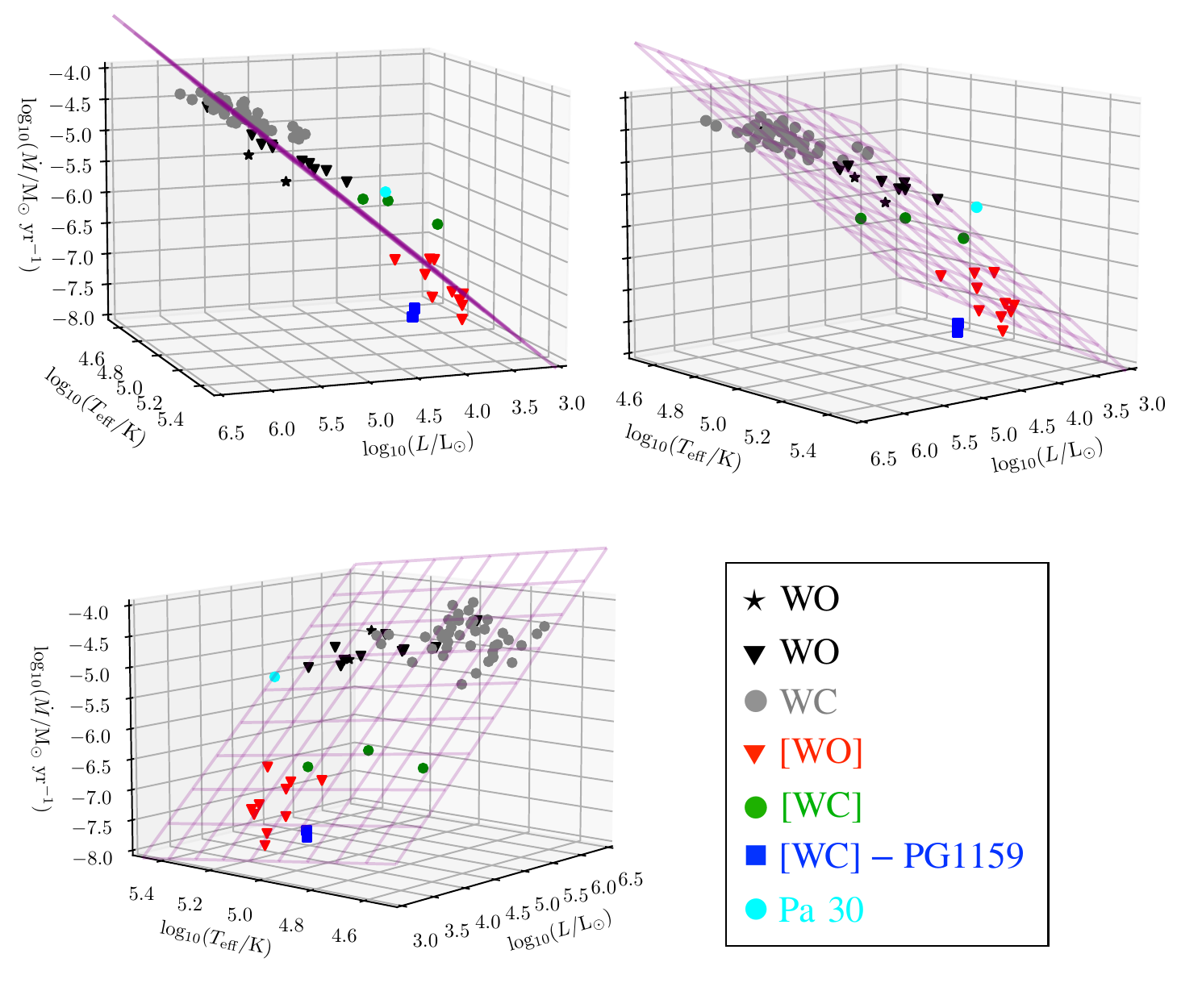}
\caption{$\dot{M}(L,T_\mathrm{eff})$ diagrams for all WR-type stars discussed in this paper. Different rows show the same diagram but through different viewing angles. The plane defined by Eq.\,\eqref{eq:plane} is overplotted in all panels with a (pink) grid.}
\label{fig:3D}
\end{center} 
\end{figure*}

One of the biggest unsolved issues in properly describing stellar evolution is the complexity and uncertainty of the mass-loss rate $\dot{M}$, which needs to be prescribed in stellar evolution models. For a given luminosity, $\dot{M}$ is known to be larger for WR-type stars than for (usually H-rich) counterparts such as OB stars or O-type CSPNe. 

We derived the wind properties of WO-type low-mass stars and for the first time demonstrated that their wind parameter behaviour aligns in various planes with results for WO-type objects evolving from massive stars. While the formulae derived above could already be used as analytical recipes to improve stellar evolutionary models, they all provide relations between just two parameters, which often limits their applicability. We now investigate if we can extend the dimensionality and potentially gain also more overarching insights into the WR-type stars in our sample.

In Fig.\,\ref{fig:3D}, we compare the fundamental stellar parameters $L$ and $T_\mathrm{eff}$ with the predictions for $\dot{M}$ from the stellar atmosphere modelling for all stars discussed in this paper in a 3D $(L, T_\mathrm{eff},\dot{M})$ diagram. Inspecting this figure through different viewing angles, we noted that all stars seem to broadly align with a plane (see Fig.\,\ref{fig:3D}) in this 3D diagram. A fit to all the WR-type stars in the ($L, T_\mathrm{eff}, \dot{M}$)-space results in 
\begin{equation}
\begin{aligned}
    \log_{10}\left(\frac{\dot{M}}{\mathrm{M}_\odot~\mathrm{yr}^{-1}}\right) = (1.23 \pm 0.42)\log_{10}\left(\frac{L}{\mathrm{L}_\odot} \right)\\ 
    - (0.68\pm0.57)\log_{10}\left(\frac{T_\mathrm{eff}}{\mathrm{K}} \right)\\ 
    - (8.11\pm0.46),
    \label{eq:plane}
\end{aligned}   
\end{equation}
\noindent were the errors correspond to the standard deviation obtained in each dimension. The plane defined by Eq.\,\eqref{eq:plane} is also plotted alongside all the stars analysed here in Fig.\,\ref{fig:3D} with selected projections of the $\dot{M}(L,T_\mathrm{eff})$ diagram. To further illustrate this, we include a GIF file as an online attachment to this article (see Appendix~\ref{app:3D}). 

Given our analytical relationships described in previous sections, which differentiate WC+[WC] from WO+[WO] stars into two distinctive populations in the $\dot{M}$--$L$ and $D_\mathrm{mom}$--$L$ space, it might be contradictory to define a single plane for all [WO], [WC], WO, and WC stars. However, separated plane fits to the two populations of [WO]+WO and [WC]+WC stars did not result in statistically different equations than the one provided by Eq.\,\eqref{eq:plane}. Moreover, the diagrams shown in Fig.~\ref{fig:wind} and Fig.~\ref{fig:trans_mass} are projections of the 3D plane defined by Eq.~(\ref{eq:plane}).

The resulting Eq.\,\eqref{eq:plane} is largely dominated by the WO+[WO] sample in the luminosity regime, which is also evident from the similarity in the $L$-dependence of Eqs.\,\eqref{eq:WO} and \eqref{eq:plane}. The massive WC stars seem to roughly align with the plane by including the $T_\text{eff}$-dependency while the biggest outliers are the low-mass [WC] stars (above the plane) and the [WC]-PG1159 stars (below the plane). The WO-type central star of Pa 30 seems to align with the plane, especially given the considerable uncertainty of several of its parameters. 

The anticorrelation with temperature in Eq.\,\eqref{eq:plane} is expected from the 2D-findings for $\dot{M}(T_\text{eff})$ above. In fact, Eq.~(\ref{eq:plane}) seems to be largely dominated by massive stars in this regime, given its very similar dependence as that of Eq.~(\ref{eq:massloss1}), which is in line with earlier studies for massive WR stars \citep[e.g.,][]{Grafener2008,Sander2023}. However, $T_\text{eff}$ defined at $\tau$=2/3 does not correspond to the location of the quasi-hydrostatic layers in many WR stars \citep[e.g.,][]{Grafener2017,Grassitelli2018}, where the mass-loss rate is actually fixed, and more direct trends in $\dot{M}$ are expected from the effective temperature at the critical point of the wind \citep{Sander2023}. The latter is inaccessible with the current data as this would require a larger dedicated dynamically-consistent modelling \citep{Grafener2005,Sander2017} for all involved targets.

\subsection{On [WC]-PG1159 stars}

The weaker spectral lines of the [WC]-PG1159 stars compared to [WC] stars mean that their weaker winds probably share some similarities with those of massive early-type WN/O stars that have a combined absorption and emission-line spectrum \citep[e.g.,][]{Morgan1991,Neugent2017}. Similar to them, the [WC]-PG1159 stars have low wind efficiencies of $\eta<1$ and correspondingly low $\dot{M}$ values. Nonetheless, they are grouped with the [WO]+WO stars in various other quantities (e.g., $T_\text{eff}$, $\dot{M}_\text{t}$, $\varv_\infty$). 
This is not too surprising since these stars exhibit very similar properties as those defined for [WO] stars. For example, the optical spectra of A\,30 and A\,78 show the clear presence of the O\,{\sc vi} at $\sim$3820~\AA\, and some other broad lines corresponding to O\,{\sc v} and O\,{\sc vi}, in addition to the classic ``blue bump'' produced by He\,{\sc ii} at 4686~\AA\, and the ``red bump'' caused by the C\,{\sc iv} at 5808~\AA\,\citep[see, e.g.,][]{Toala2015}. 
One might even argue that [WC]-PG1159 stars complete the $\eta$--$\dot{M}_\mathrm{t}$ sequence, but given their small number we refrain from adding them to any of the linear regression approaches. If that would be the case, the slope in Eq.~(\ref{eq:mt}) might be a bit steeper.

Other objects that have been suggested to have experienced a born-again event are A\,58 and HuBi\,1. They have been estimated to have experienced a VLTP a few hundred years ago \citep[see][]{Clayton1997,Guerrero2018} while A\,30 and A\,78 experienced it about 1000\,yr ago. However, A\,58 and HuBi\,1 have not been included in our analysis due to their questionable distances. Particularly, we note that HuBi\,1 has been classified as a late [WC10] object with a stellar temperature of 38~kK, but the evolution after the born-again event is expected to take the CSPN into the high $T_\mathrm{eff}$ range in a short period of time \citep[see][and references therein]{MB2006}, that is changing its stellar and wind properties dramatically, going from $\eta > 1$ to $\eta < 1$. One might expect that, at least for born-again objects, their CSPN to go from the [WC]+WC sequences to those defined for [WO]+WO in short times. Of course, this might be the case for all [WR]-type CSPNe but such assertions can only be shown by future stellar evolution models for H-deficient stars given that different evolutionary paths might depend on other parameters such as initial mass, metallicity, and rotation.  

Finally, we note that the CSPN of NGC\,2371 has also been recently classified as a [WC]-PG1159 transition object by \citet{Corsico2021} without giving strong arguments. It is true that this star exhibits the C~{\sc iv} at $\sim$4440~\AA\, in absorption (instead of in emission), but it does not exhibit the typical He~{\sc ii} 4539~\AA\, and C~{\sc iv} 4552~\AA\, absorption features observed in [WC]-PG1159 stars. In this work we showed that the CSPN of NGC\,2371 and PC\,22 show some similar properties as the [WC]-PGG1159 objects A\,30 and A\,78. This situation seem to suggest that NGC\,2371 and PC\,22 might be at the verge of becoming [WC]-PG1159 stars, which consequently means that there are different evolutionary paths to create [WC]-PG1159 stars. Only future stellar evolution models will help peering into this interesting result.

\section{Summary}
\label{sec:summary}

We presented an analysis of the stellar and wind properties of low-mass [WO] and [WC] stars in conjunction with their massive counterparts (WO and WC). Including these [WR]-type stars in our analysis helped us peer into the low-luminosity range ($\log_{10}(L/$L$_\odot) <$4.5) of the WR phenomenon. We identify different regimes depending on the spectral type of the C and O sequences. In addition, we found general trends followed by both massive and low-mass WR-type stars that seem to describe in general terms the WR phenomenon regardless of mass. 

Our main findings can be summarised as follows:
\begin{itemize}

\item We demonstrated that [WC] stars complement sequences previously defined for massive WC stars regardless of the sub-type. Linear regression fits predict dependencies of $\dot{M} \propto L^{0.88\pm0.03}$ and $D_\mathrm{mom} \propto L^{1.21\pm0.16}$ for [WC]+WC stars. The former is in line with what was initially predicted by \citet{NugisLamers2000} of $\dot{M} \propto L^{0.84}$ for WC stars, confirming that indeed [WC] and WC stars share similar wind properties.

\item Including $[$WO$]$ stars in our analysis help us to predict wind properties for WO stars for the first time. [WO] stars naturally align with massive WO stars and form a different sequence than those defined by [WC]+WC stars. However, our predictions for [WO]+WO stars are not definite given the inconsistencies between different groups that have analysed WO stars. Nevertheless, our samples indicate that the slopes in the relationships for [WO]+WO stars are steeper than those for [WC]+WC stars.

\item Stars align into two different sequences in the $\dot{M}$--$T_\mathrm{eff}$ space. For massive stars, this suggest a $\dot{M} \propto T_\mathrm{eff}^{-0.63\pm0.15}$, while for [WR] stars this relationship resulted in $\propto T_\mathrm{eff}^{-1.88\pm0.63}$. It might be interesting to increase the number of [WC] stars to improve our statistics of low-mass [WR] stars to check if both sequences could have similar dependencies (or not).

\item Stars considered to have experienced a born-again event (a very late thermal pulse) behave very very similarly to those of the [WO]-type, but show lower wind efficiencies ($\eta < 1$).

\item We found most of our [WR] stars to be located in the optically-thin wind region in the $\varv_{\infty}$--$T_\mathrm{eff}$ diagram, with the only exceptions being those of the [WC] type. 
Yet, there is no breakdown of the $\dot{M}$-$L$ trend seen in our sample of WR and [WR] stars, contrary to the  ``weak wind phenomenon'' observed for massive O-type stars.

\item Using the concept of the transition mass-loss rate $\dot{M}_\mathrm{trans}$, we confirm the empirically determined mass-loss rates for [WO] stars at the single-scattering limit, underlining the robustness of our relations. We further find first evidence that the difference between wind efficiency and flux-weighted optical depth at the sonic point seems to be different between low- and high-mass WR stars.

\item By using all stars ([WR]+WR), we disclosed general trends for the WR phenomenon. They all align into a single sequence in the $\dot{M}_\mathrm{t}$--$T_\mathrm{eff}$ space with $\dot{M}_\mathrm{t} \propto T_\mathrm{eff}^{-1.95\pm0.16}$, which is very close to the predictions of $\dot{M}_\mathrm{t} \propto T_\mathrm{eff}^{-2.04\pm0.04}$ by \citet{Sander2023} only for massive stars, which suggest a universal scaling law of $\dot{M}_\mathrm{t} \propto T_\mathrm{eff}^{-2}$. Another general trend was unveiled in the $\eta$--$\dot{M}_\mathrm{t}$ space, with $\eta \propto \dot{M}_\mathrm{t}^{0.43\pm0.06}$. These two relationships seem to suggest that $\dot{M}_\mathrm{t}$ might be the best way to compare the two apparently disconnected populations of massive WR and low-mass [WR] stars.

\item Finally, we noticed that all WR-type stars analysed here seem to broadly align within a fundamental plane defined in the $(L, T_\mathrm{eff}, \dot{M})$ space, where $\dot{M} \propto L^{-1.23\pm0.42} T_\mathrm{eff}^{-0.68\pm0.57}$. The plane is dominated by the [WO]+WO sample in the $L$-regime, but mostly dominated by massive stars in the $T_\mathrm{eff}$-regime. Even stars with an unusual history such as A\,30, A\,78 and Pa 30 do not deviate significantly from this plane. It needs to be seen whether objects such as the [WC] stars and their massive-star counterparts could form another plane or mark an extension of the plane that is influence by additional parameters such as abundance, but for this further studies with more objects are required.

\end{itemize}

We note that in order to corroborate our findings, a larger sample of [WR]-type stars with reliable distance estimates and spectral analyses with state-of-the-art stellar atmosphere models of high-quality spectra, in particular UV spectra, is needed. In addition, a coherent analysis of the stellar and wind parameters of the complete sample of the -- not so numerous -- massive WO stars would be most desirable. 
Finally, testing whether hydrogen-depleted WR stars of the N-sequence (WN and [WN/WC] stars) align with our findings will mark another step into understanding the WR phenomenon across high and low-mass stars.

\section*{Acknowledgements} 

The authors thank an anonymous referee for comments and suggestions that improve our manuscript. The results presented in this work were sparked by a discussion during the conference ``The Wolf-Rayet phenomenon in the Universe'' celebrated in June 2023 in Morelia, Mexico. JAT thanks support from the UNAM PAPIIT project IN102324. JAT is grateful to DRC for their support during the past years that resulted in the present work. AACS is supported by the Deutsche Forschungsgemeinschaft (DFG - German Research Foundation) in the form of an Emmy Noether Research Group -- Project-ID 445674056 (SA4064/1-1, PI Sander). AACS is further supported by funding from the Federal Ministry of Education and Research (BMBF) and the Baden-Württemberg Ministry of Science as part of the Excellence Strategy of the German Federal and State Governments. This paper has benefited from discussions
at the International Space Science Institute (ISSI) in Bern, through ISSI International Team project 512 (Multiwavelength View on Massive Stars in the Era of Multimessenger Astronomy, PI Oskinova). This work has made a large use of NASA’s Astrophysics Data System (ADS). 
 




\bibliographystyle{mn2e}
\bibliography{wr-winds}

\section*{DATA AVAILABILITY}

The data underlying this article will be shared on reasonable request
to the corresponding author.

\appendix

\section{Fundamental plane of WR-type stars}
\label{app:3D}

Here we include a GIF animation of the fundamental plane defined by all WO- and WC-type stars analysed here. The symbol and colour code of the GIF file is the same as that of Fig.~\ref{fig:3D}.



\bsp	
\label{lastpage}
\end{document}